%% file: raja_isvhecri06.tex
\documentclass[fleqn,twoside]{article}
\usepackage[headings]{espcrc2}

\readRCS
$Id: espcrc2.tex,v 1.2 2004/02/24 11:22:11 spepping Exp $
\usepackage{graphicx}
\usepackage[figuresright]{rotating}

\newcommand{\AmS}{{\protect\the\textfont2
  A\kern-.1667em\lower.5ex\hbox{M}\kern-.125emS}}

\title{The Main Injector Particle Physics Experiment (MIPP FNAL E-907) 
at Fermilab - status and plans}

\author{Rajendran Raja\address{
        Fermilab, P.O. Box 500, Batavia, ILlinois 60510, U.S.A}}
       
\runtitle{The MIPP Experiment at Fermilab}
\runauthor{R.Raja}

\begin{document}

\begin{abstract}
We describe the status of the Main Injector particle production
Experiment (MIPP) at Fermilab which has to date acquired 18~million
events of particle interactions using (5~GeV/c-120~GeV/c) $\pi^\pm,
K^\pm$ and $p^\pm$ beams on various targets. We describe plans to
upgrade the data acquisition speed of MIPP to make it run 100 times
faster which will enable us to obtain particle production data of
unprecdented quality and statistics on a wide variety of nuclear
targets including nitrogen which is of importance to cosmic ray
physics.
\vspace{1pc}
\end{abstract}
\maketitle
\input mipp_status
\input atmospheric
\input run_plan

\end{document}

%% file: mipp_status.tex
\section{Current Status of the MIPP Experiment}

We describe status report on the MIPP~\cite{mipp} experiment and its
performance to date. MIPP is situated in the Meson Center
beamline at Fermilab. It received  approval~\cite{proposal} in
November 2001 and has installed and operated  both the experiment and a newly
designed secondary beamline in the interim. It received its first
beams in March 2004, had an engineering run to commission the detector
in 2004 and had its physics data-taking run in the period January
2005-March 2006. The experiment is currently busy analyzing its data.

\begin{figure}[htb!]
\includegraphics[width=0.5\textwidth]{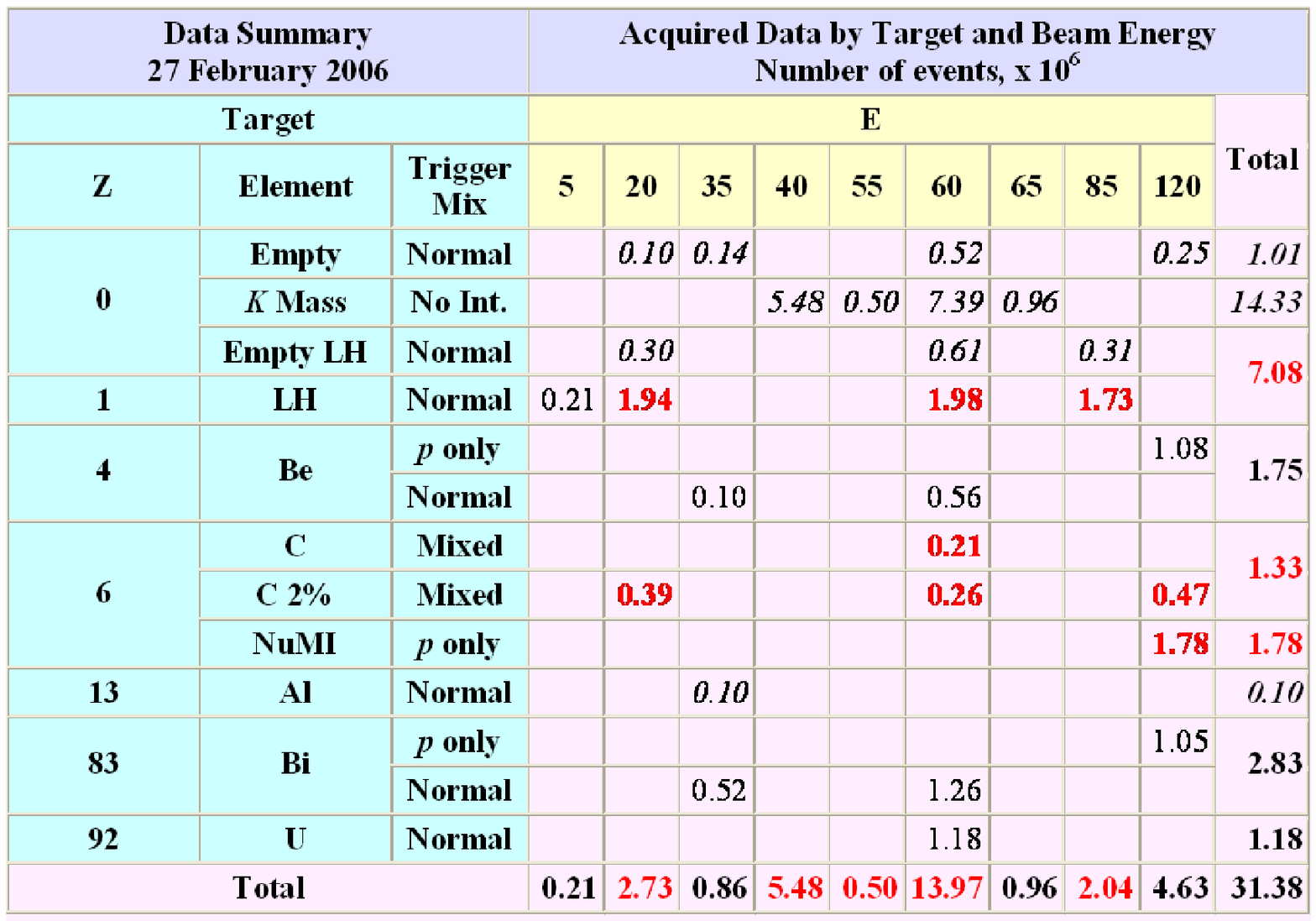}
\caption{The data taken during the first MIPP run as a function of nucleus. 
The numbers are in millions of events. During the last month of the
run, the Jolly Green Giant magnet coils developed shorts. This time
was used to acquire data without the TPC for exploring the feasibility
of measuring the charged kaon mass using the RICH radii.}
\label{tab1}
\end{figure}

MIPP is designed primarily as an experiment to measure and study in
detail the dynamics associated with non-perturbative strong
interactions. It has nearly 100\% acceptance for charged particles and
excellent momentum resolution.  Using particle identification
techniques that encompass $dE/dx$, time-of-flight~\cite{tof},
Multi-Cell \v Cerenkov~\cite{e690} and a Ring Imaging \v Cerenkov
(RICH)detector~\cite{rich}, MIPP is designed to identify charged particles
at the 3$\sigma$ or better level in nearly all of its final state
phase space.  MIPP has acquired data of unparalleled quality and
statistics for beam momenta ranging from 5~GeV/c to 90~GeV/c for 6
beam species ($\pi^\pm, K^\pm~ $and$~ p^\pm$) on a variety of targets
as shown in Figure~\ref{tab1}.

 An important aspect of MIPP data-taking was the measurement of
 particle production off the NuMI~\cite{minos} target in order to
 minimize the systematics in the near/far detector ratio in the
 MINOS~\cite{minos} experiment.  MIPP also made measurements with
 proton beams off various nuclei for the needs of proton
 radiography~\cite{proposal}.

Another physics motivation behind MIPP is to restart the study of
non-perturbative QCD interactions. Currently available data are of poor
quality, and sparsely populate the beam momentum, $p_T$, and 
atomic weight phase space
that makes comparisons between different experiments difficult. The MIPP
TPC~\cite{tpc} digitizes the
charged tracks in three dimensions, obviating the need for track
matching across stereo views.  Coupled with the particle
identification capability of MIPP, the data from MIPP would add
significantly to our knowledge base of non-perturbative QCD. This
would help test inclusive scaling relations and also scaling nuclear
reactions.

\subsection{Experimental Setup}

We designed a secondary beam~\cite{carol} specific to our needs. 
The resonantly extracted protons from the Fermilab Main Injector 
are transported down the Meson Center line. 
They impinge on a 20~cm long copper target producing
secondary beam particles. This target is imaged onto an adjustable
momentum selection collimator which controls the momentum spread of
the beam. This collimator is re-imaged on to our interaction target
placed next to the TPC. The beam is tracked using three beam chambers
and identified using two differential \v Cerenkovs~\cite{bckov} filled
with gas, the composition and the pressure of which can be varied
within limits depending on the beam momentum and charge.

\begin{figure}[htb!]
\includegraphics[width=0.5\textwidth]{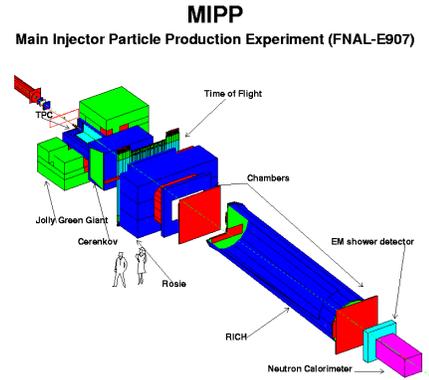}
\caption{The experimental setup. The picture is a rendition in 
Geant3, which is used to simulate the detector.}
\label{mipp}
\end{figure}

Figure~\ref{mipp} shows the layout of the apparatus. The TPC sits in a
wide aperture magnet (the Jolly Green Giant) which has a peak field of
0.7 tesla. Downstream of the TPC are a 96 mirror multi-cell \v
Cerenkov detector filled with $C_4F_{10}$ gas, and a time of flight
system. This is followed by a large aperture magnet (Rosie) which runs
in opposite polarity (at -0.6 tesla) to the Jolly Green Giant to bend
the particles back into the Ring Imaging \v Cerenkov counter. The RICH
has $CO_2$ as the radiator and an array of phototubes of 32 rows and
89 columns~\cite{fire}.
Downstream of the RICH we have an electromagnetic
calorimeter~\cite{ecal}and a hadron
calorimeter~\cite{hcal} to
measure forward-going photons and neutrons. The electromagnetic
calorimeter provides a means of distinguishing forward neutrons 
from photons and will also serve as a device to measure the electron
content of our beam at lower energies, which will be useful for
measuring cross sections.

\begin{figure}[hbtp] 
\begin{center}
\begin{minipage}{15pc}
\includegraphics[width=0.8\textwidth]{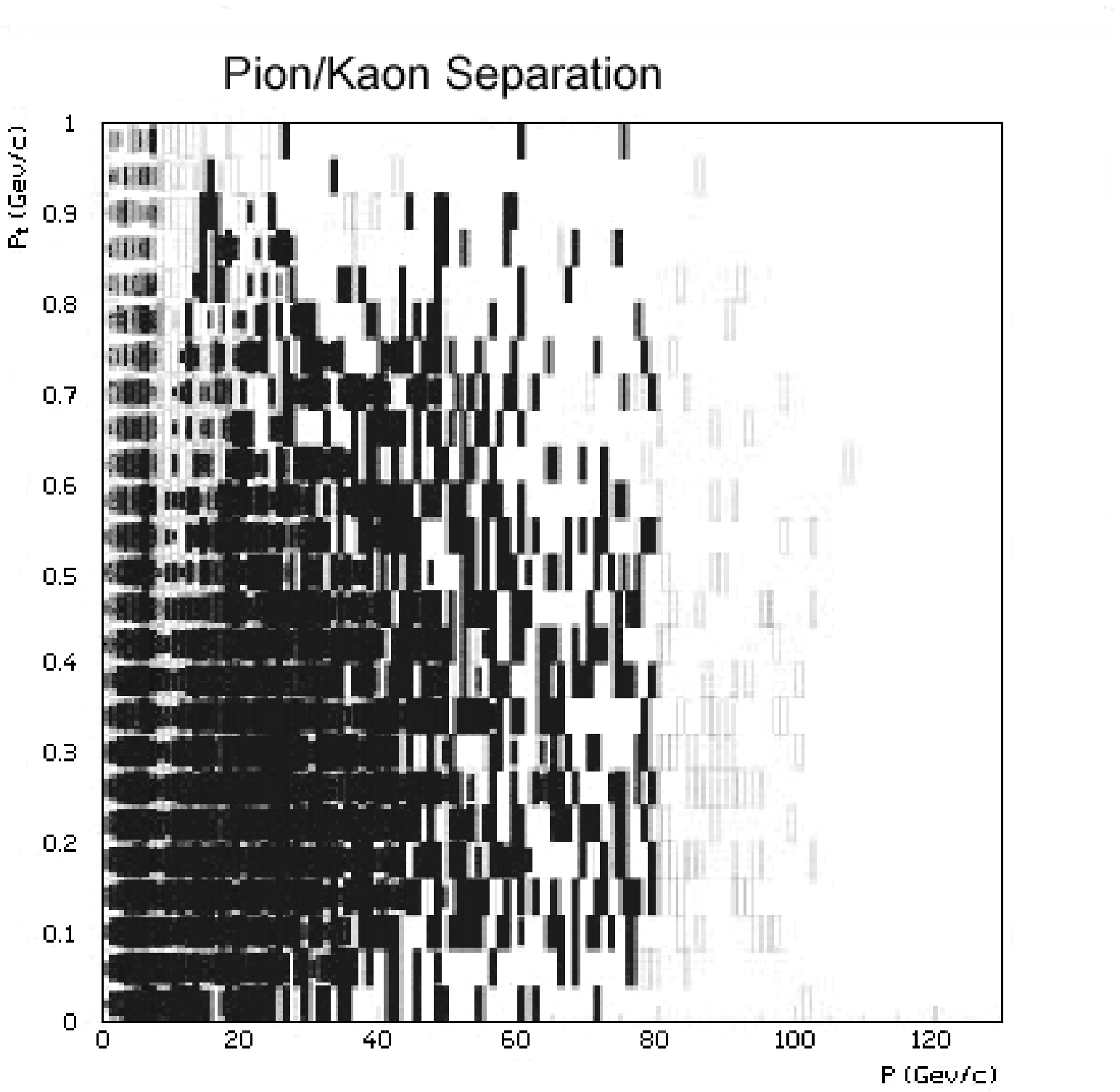}
\end{minipage}
\end{center}
\begin{center}
\begin{minipage}{15pc}
\includegraphics[width=0.8\textwidth]{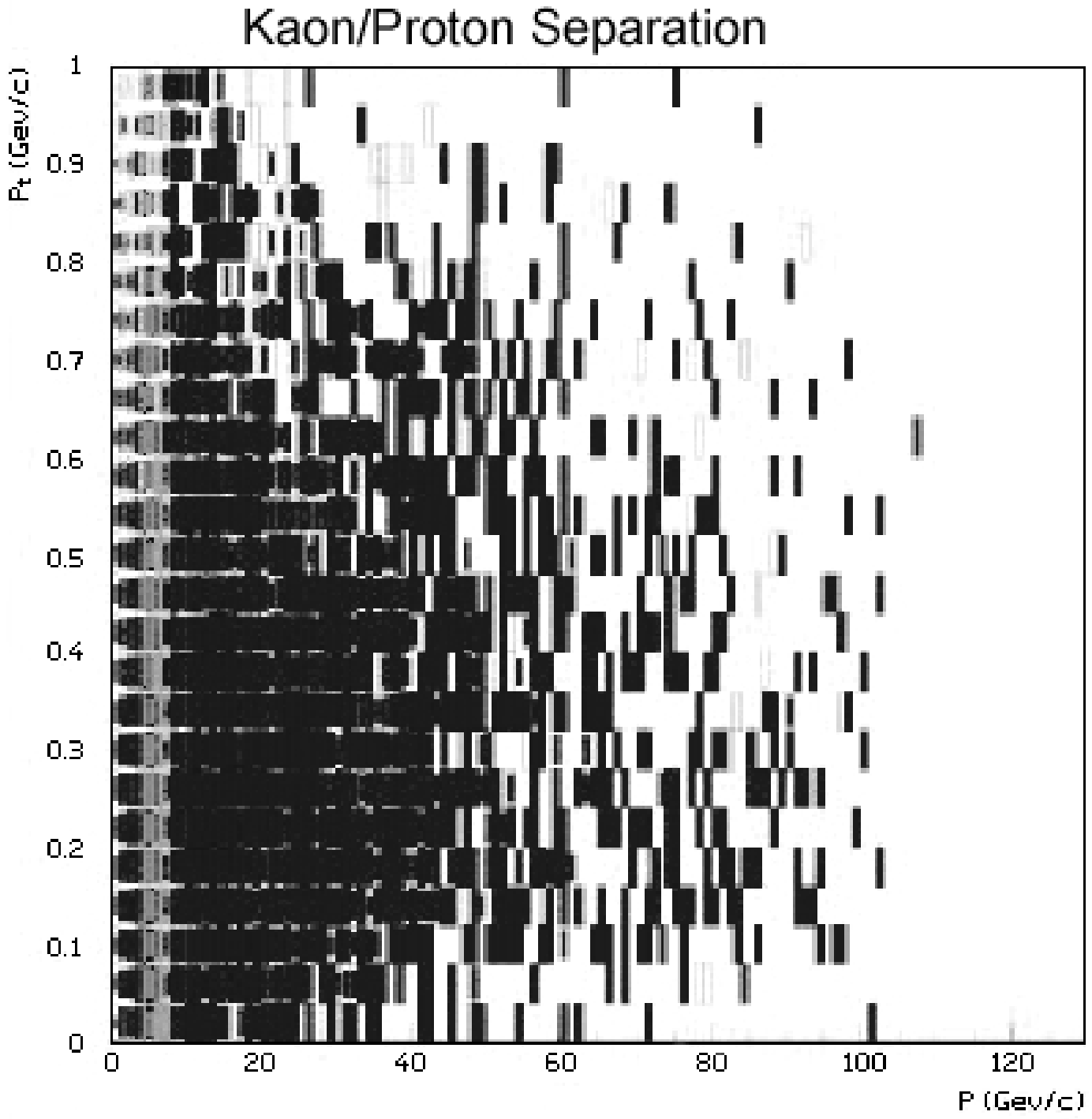}
\end{minipage}
\end{center}
\caption{Particle identification plots for pion/kaon separation and for
kaon/proton separation as a function of the longitudinal and
transverse momentum of the outgoing final state particle. Black
indicates separation at the $3\sigma$ level or better and grey
indicates separation at the $1-3\sigma$ level. The boxes at largest
values of the longitudinal momenta suffer from lack of kaon
statistics.}
\label{pid}
\end{figure}

MIPP uses $dE/dx$ in the TPC to separate pions, kaons and protons for
momenta less than $\approx$ 1~GeV/c  and the time of flight array of
counters to do the particle identification for momenta less than
2~GeV/c. The multi-cell \v Cerenkov detector~\cite{e690} contributes
to particle identification in the momentum range $\approx$ 2.5~GeV/c-14~GeV/c
and the RICH~\cite{rich} for momenta higher than this. By combining
information from all counters, we get the expected particle
identification separation for $K/p$ and $\pi/K$ as shown in
Figure~\ref{pid}. It can be seen that excellent separation at the
$3\sigma$ or higher level exists for both $K/p$ and $\pi/p$ over
almost all of phase space.  Tracking of the beam particles and
secondary beam particles is accomplished by a set of drift
chambers~\cite{chambers1} and proportional chambers~\cite{chambers2}
each of which have 4 stereo layers.
\subsection{Some results from Acquired data}
\begin{figure}[htb!]
\begin{minipage}{15pc}
\includegraphics[width=\textwidth]{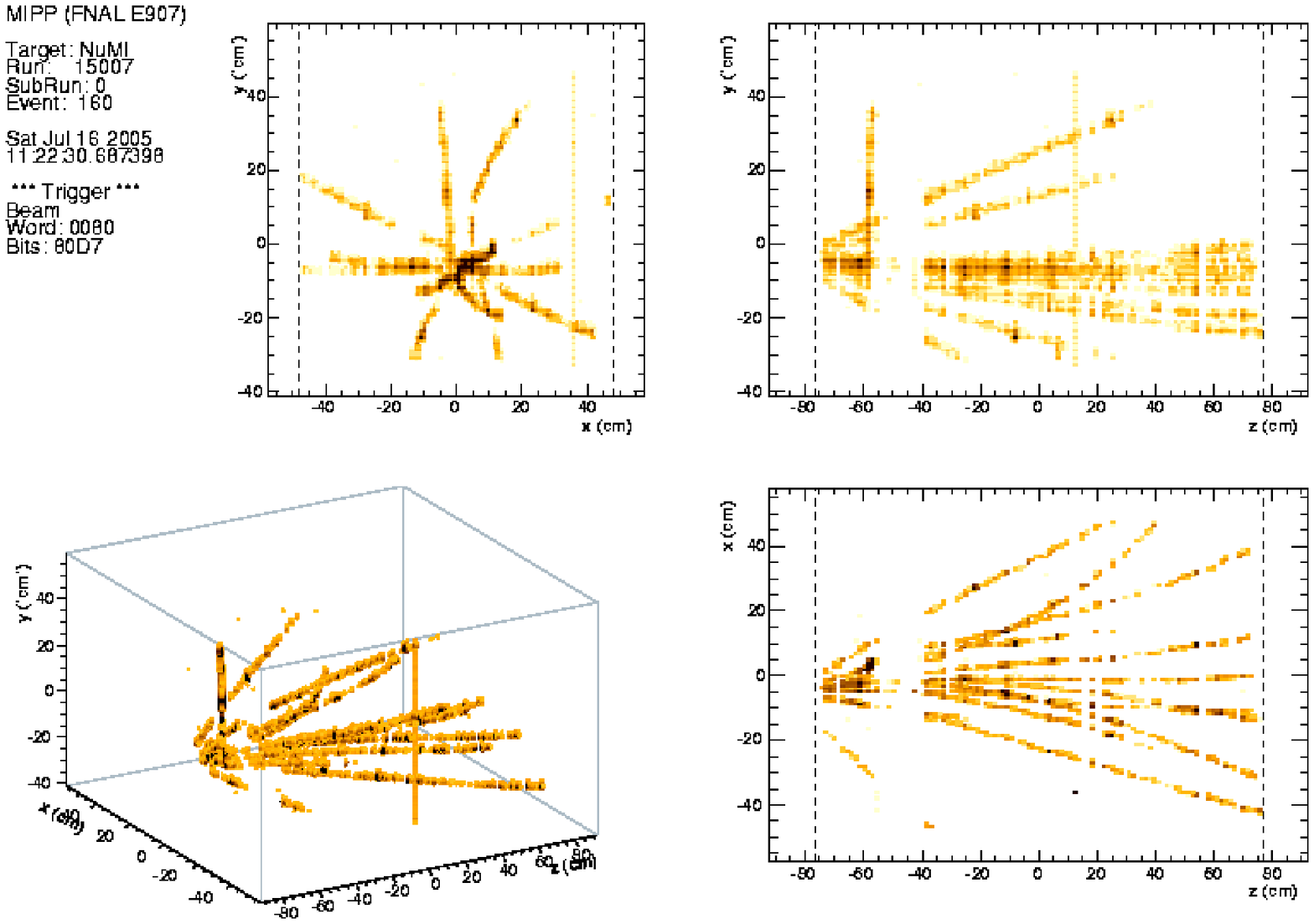}
\end{minipage}
\begin{minipage}{15pc}
\includegraphics[width=\textwidth]{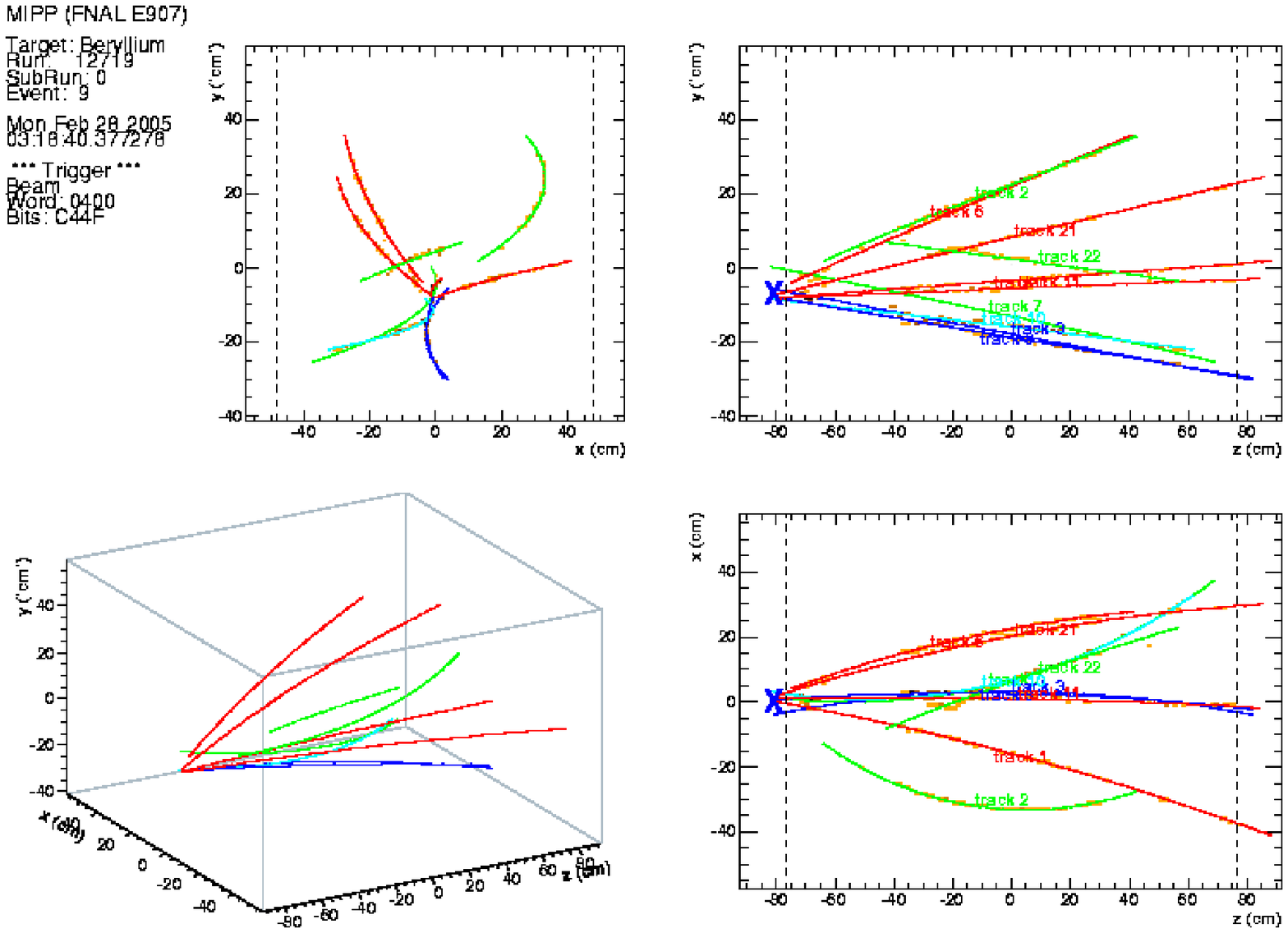}
\end{minipage}
\caption{RAW and Reconstructed TPC tracks  from two different events.}
\label{tpc1}
\end{figure}

Figure~\ref{tpc1} shows the pictures of reconstructed tracks in the
TPC obtained during the data-taking run. The tracks are digitized and
fitted as helices in three dimensions. Extrapolating three dimensional
tracks to the other chambers makes the pattern recognition
particularly easy.

\begin{figure}[htb!]
\begin{minipage}{15pc}
\includegraphics[width=\textwidth]{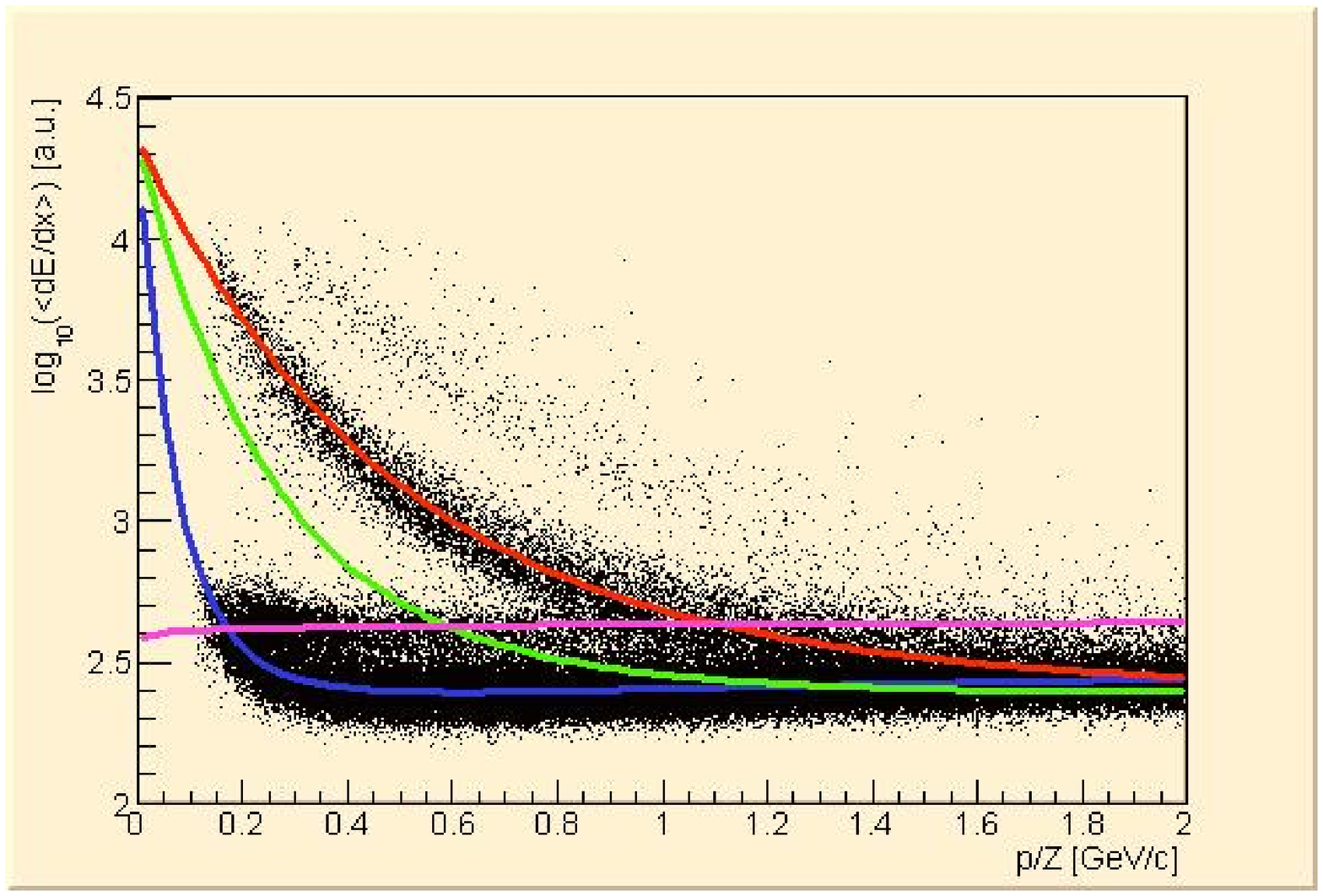}
\end{minipage}
\begin{minipage}{15pc}
\includegraphics[width=\textwidth]{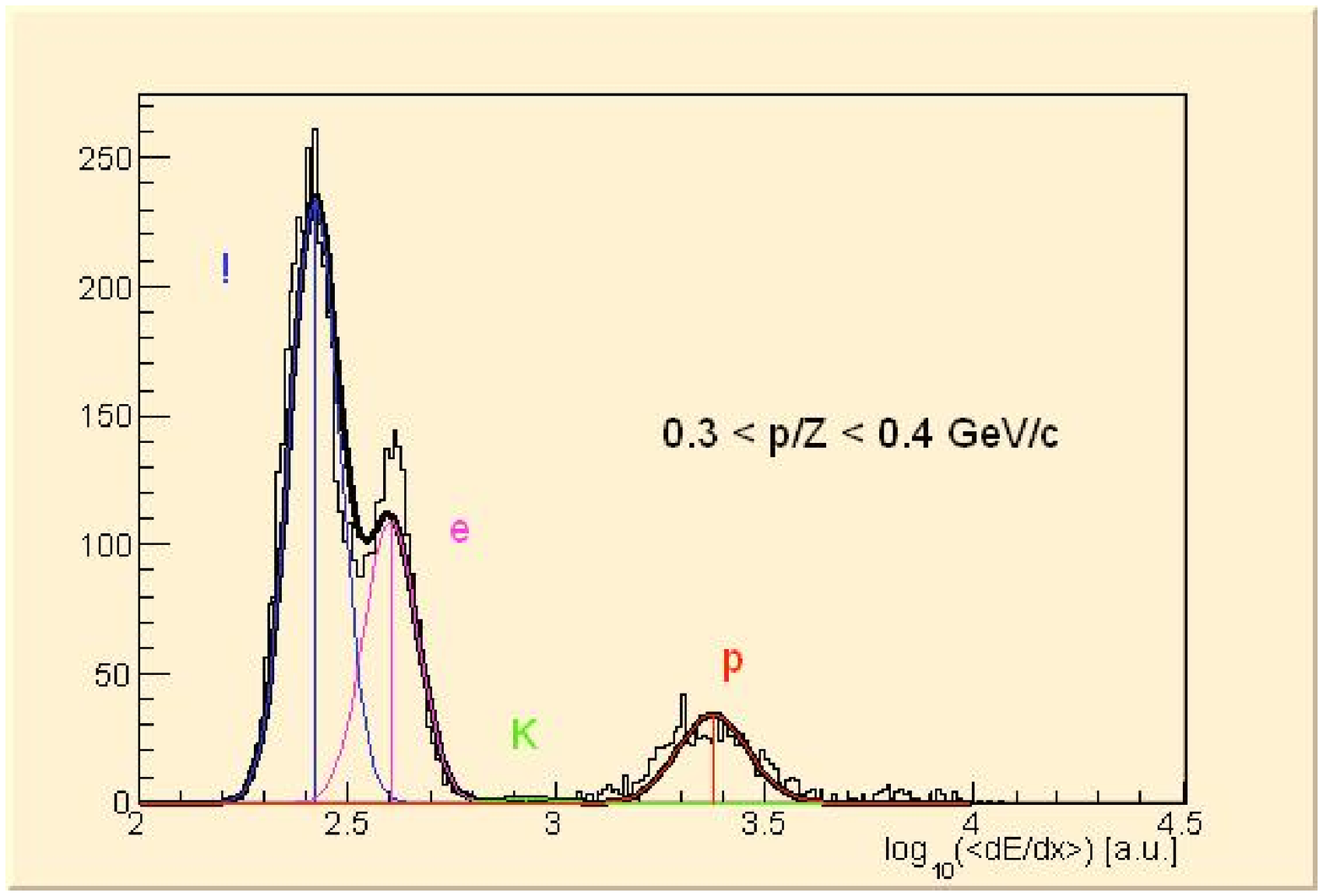}
\end{minipage}
\caption{Preliminary dE/dx distributions in the TPC The scatter plot
shows the electron, 
pion, kaon and proton peaks in the distribution as a function of
the lab momentum  for p-Carbon data. 
The second plot is the projection on the dE/dx axis
for a momentum slice 0.3~GeV/c to 0.4~GeV/c.}
\label{dedx}
\end{figure}

\begin{figure}[htb!]
\begin{minipage}{15pc}
\includegraphics[width=\textwidth]{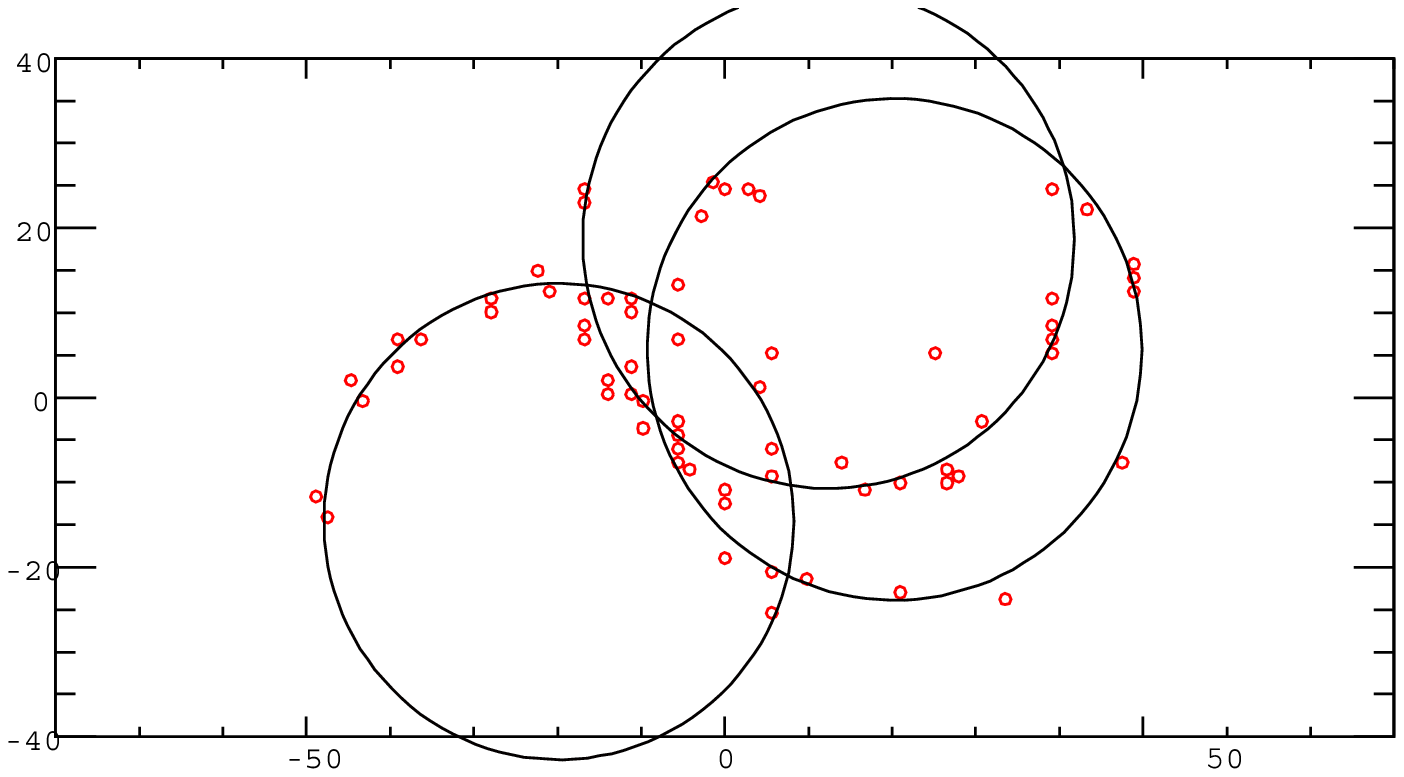}
\end{minipage}
\begin{minipage}{15pc}
\includegraphics[width=\textwidth]{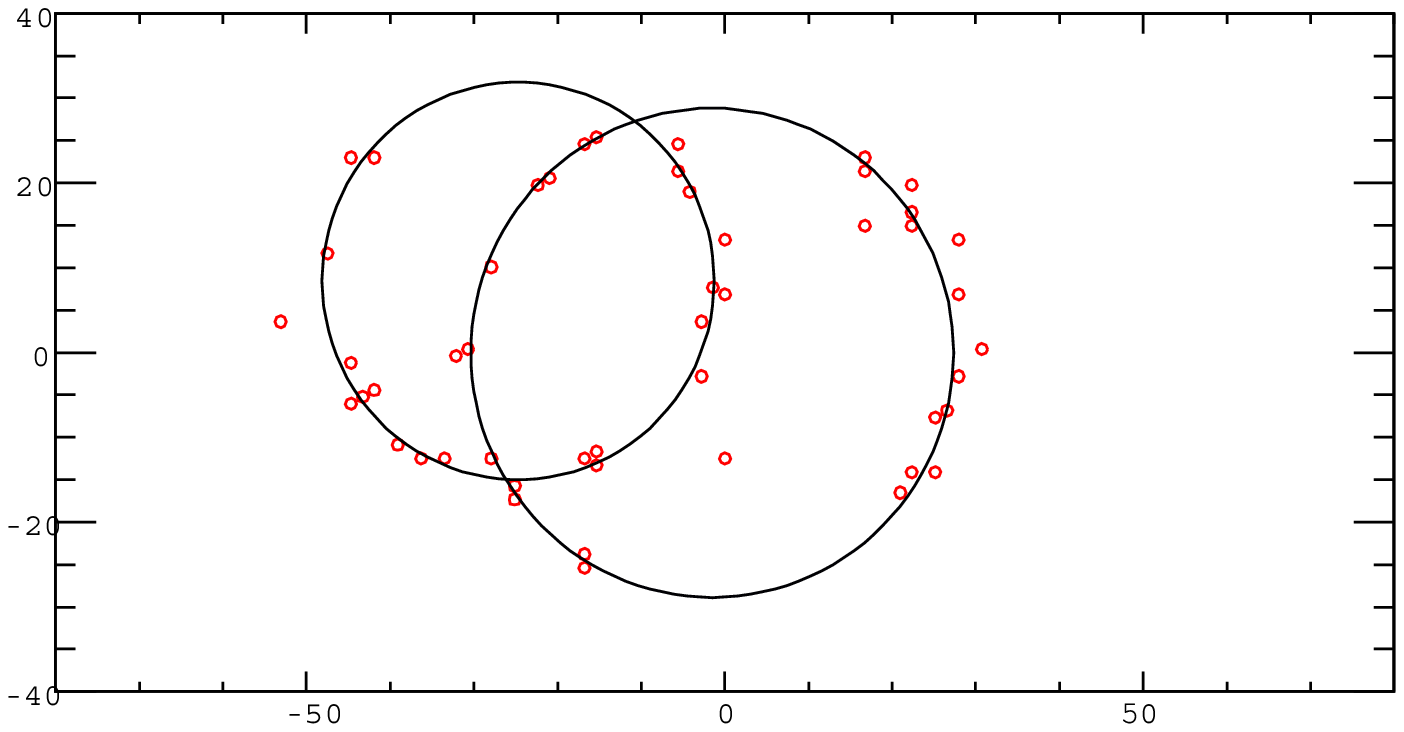}
\end{minipage}
\caption{\label{rings}Examples of events with rings in the RICH counter 
for a 40~GeV/c beam. The x and y axes are in cm.}
\end{figure}

\begin{figure}[htb!]
\centerline{
\begin{minipage}{10pc}
\includegraphics[width=\textwidth]{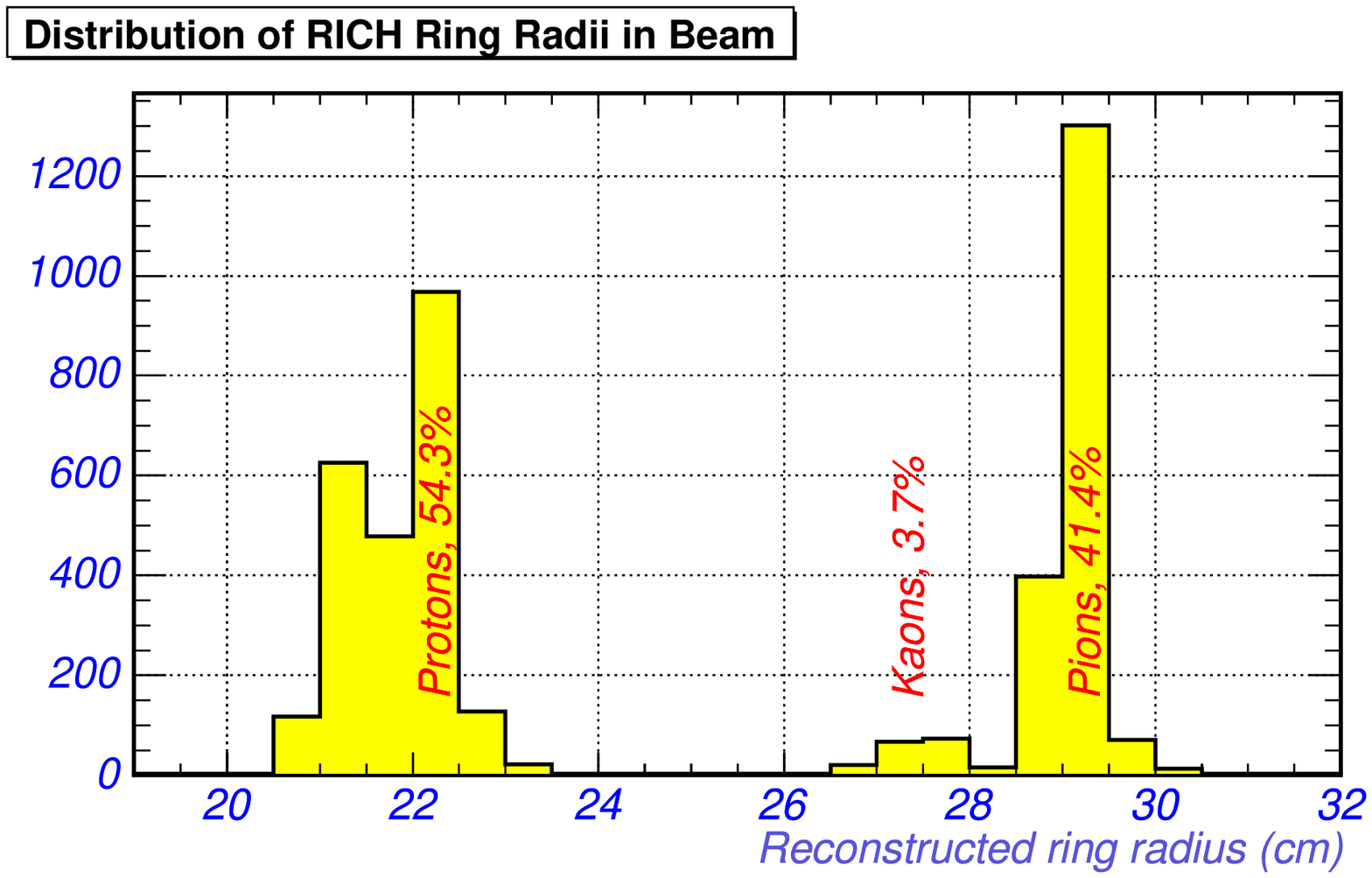}
\end{minipage}
\begin{minipage}{10pc}
\includegraphics[width=\textwidth]{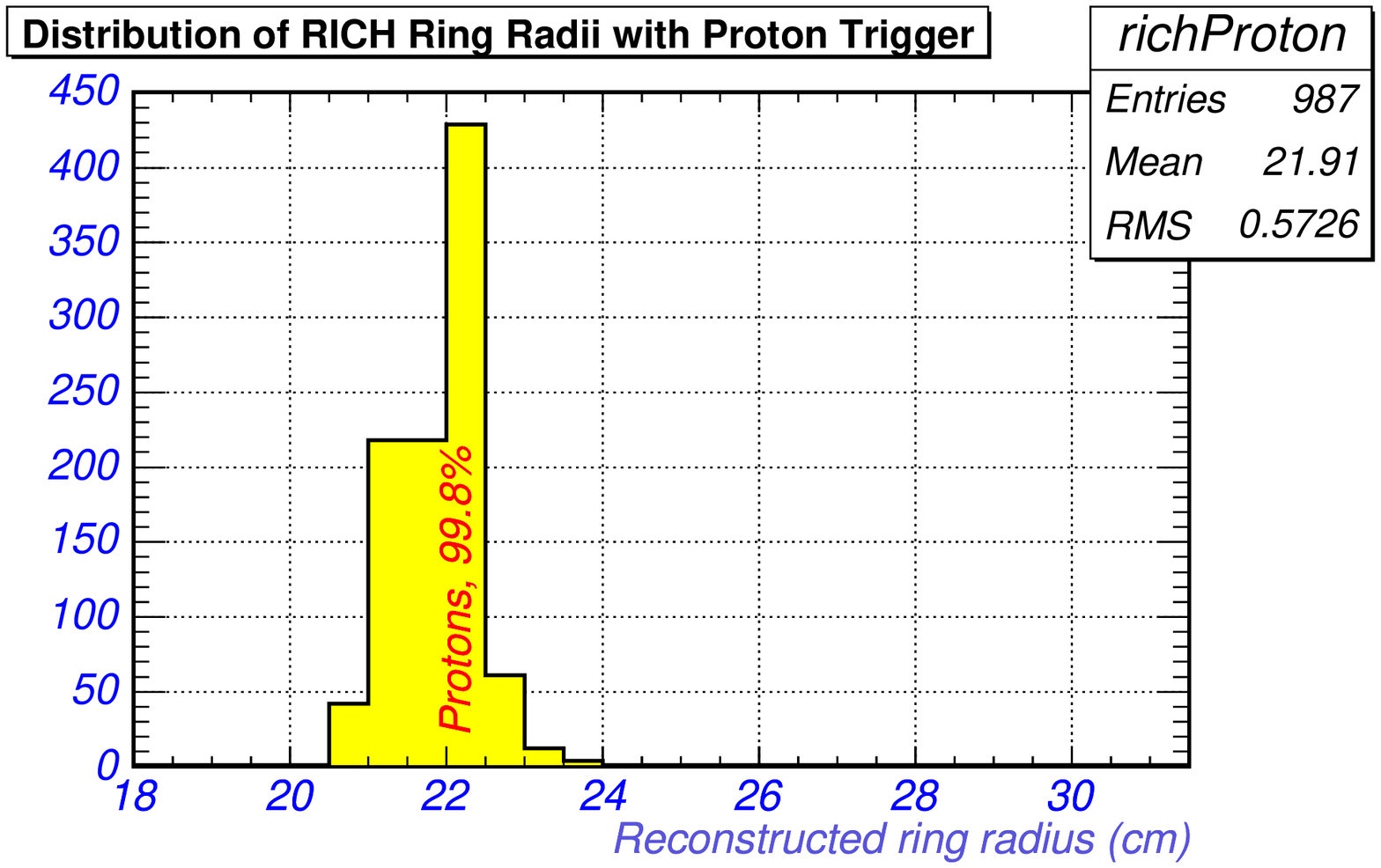}
\end{minipage}
}
\centerline{
\begin{minipage}{10pc}
\includegraphics[width=\textwidth]{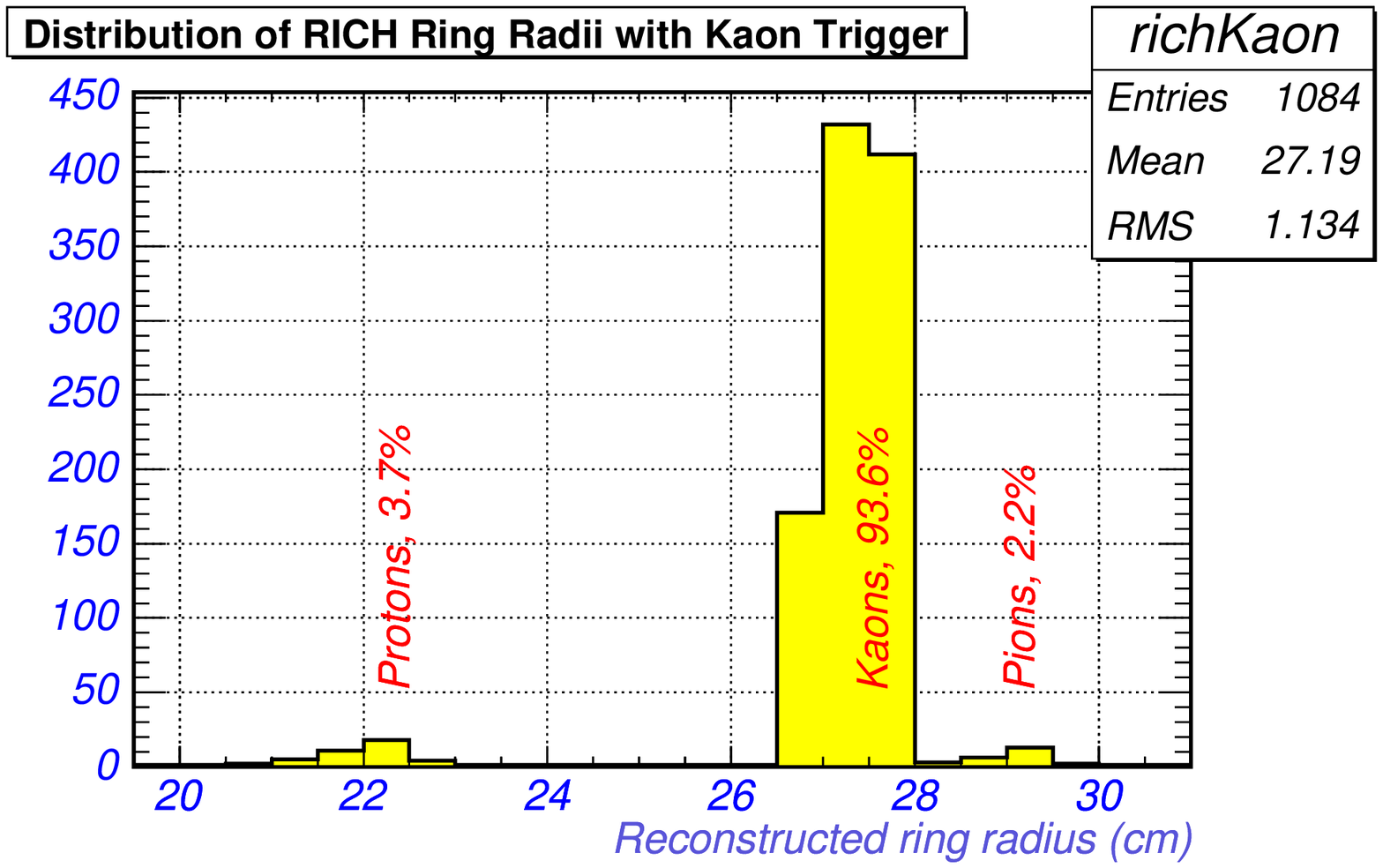}
\end{minipage}
\begin{minipage}{10pc}
\includegraphics[width=\textwidth]{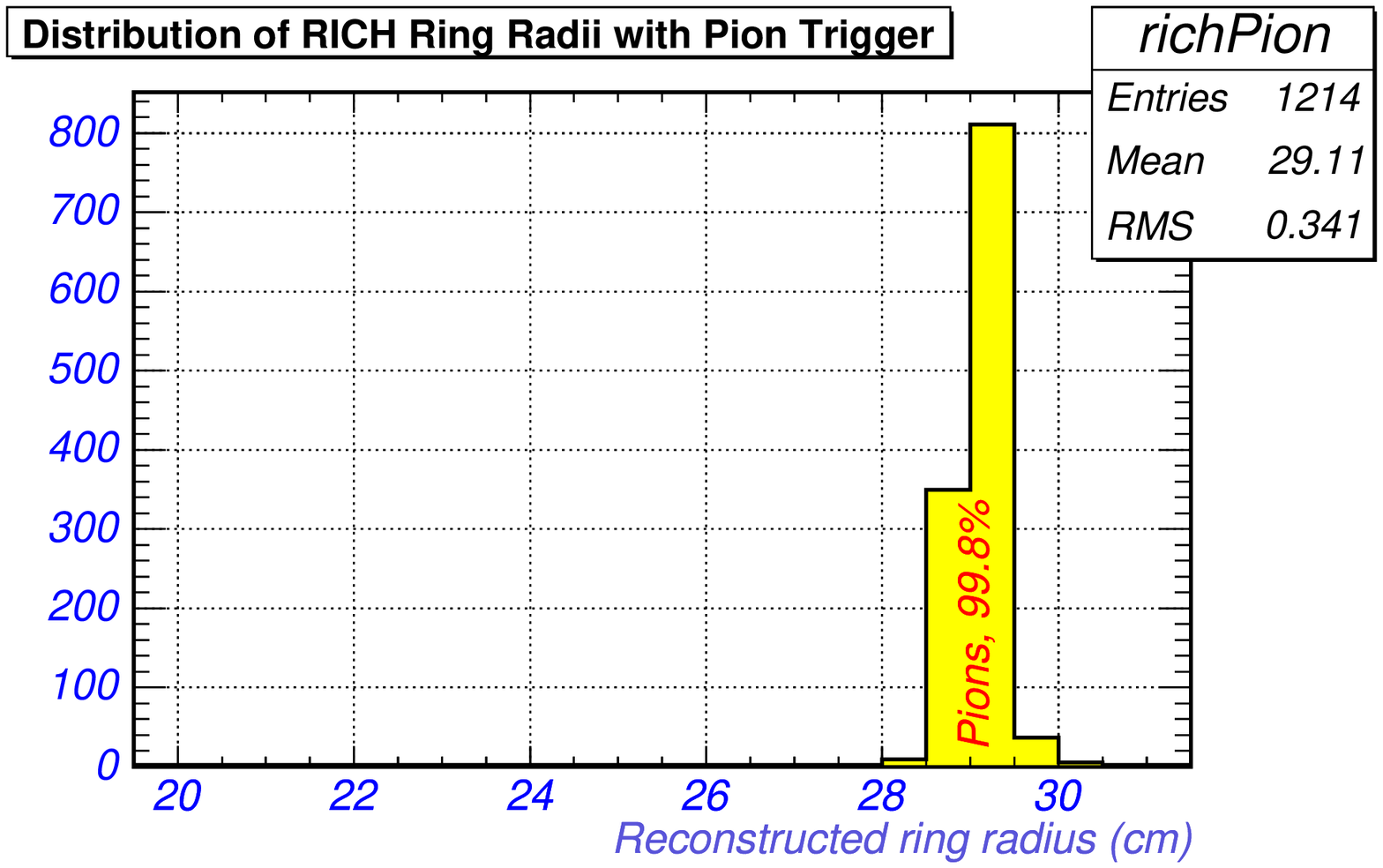}
\end{minipage}
}
\caption{An example of a 40 GeV/c primary beam (non-interacting)
trigger. The RICH identifies protons, kaons and pions by the ring
radii. The beam \v Cerenkov detectors can be used to do the same. When
the beam \v Cerenkov identification is used, one gets a very clean
separation of pions, kaons and protons in the RICH.}
\label{rbck}
\end{figure}

Figure~\ref{dedx} shows the distribution of $dE/dx$ of tracks measured
in the TPC as a function of the track momentum in a preliminary
analysis of p-Carbon data. The TPC is capable of separating pions,
protons and kaons in the momentum range below $\approx$ 1~GeV/c.
Figure~\ref{rings} shows events with multiple rings in the RICH counter. 
Figure~\ref{rbck} shows the histogram of ring radii for a +40~GeV
secondary beam. There is clean separation between pions, kaons and
protons and their relative abundances~\cite{malensek} match
expectations. Applying the particle identification trigger from the
beam \v Cerenkovs enables us to separate the three particle species
cleanly. The kaons which form ~4\% of the beam are cleanly picked out
by the beam \v Cerenkov with very simple selection criteria. These can
be made much more stringent with offline cuts to produce a very clean
kaon beam. 

The ring radius of the particle contains information on the mass of
the particle. The pion and proton masses are very well known. The
charged kaon mass, however, currently has measurement uncertainties of
the order of 60~keV. Improving the precision of both charged kaon
masses will pay dividends in rare K decay  experiments involving
charged kaons where the matrix elements depend on the kaon mass raised
to large powers. Towards the end of our physics run, when the Jolly
Green Giant magnet coils failed, we switched off the TPC and acquired
data at the rate of 300~Hz to investigate how well we can measure the
charged kaon mass. These events, whose statistics are indicated in
Figure~\ref{tab1}, are currently being analyzed to evaluate the
systematics involved in such a measurement.
\subsubsection{NuMI target measurements}
\begin{figure}[htb!]
\centerline{
\begin{minipage}{15pc}
\includegraphics[width=0.8\textwidth]{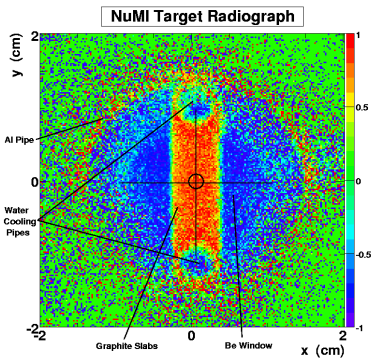}
\end{minipage}
}
\caption{Radiograph of the MINOS target. The beam direction is perpendicular 
to the paper.}
\label{minos}
\end{figure}

\begin{figure}[htb!]
\centerline{
\begin{minipage}{15pc}
\includegraphics[width=0.8\textwidth]{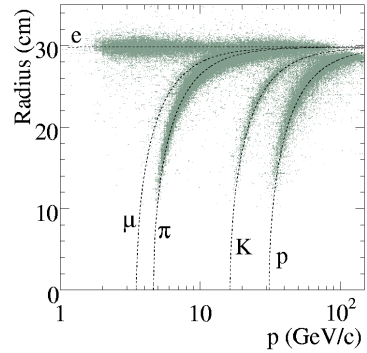}
\end{minipage}
}
\caption{Preliminary data of RICH ring radii of positive 
tracks from the NuMI target vs momentum. Superimposed are the expected curves
for $e,~\mu,~K$ and $p$ particles.}

\label{ringrads}
\end{figure}

MIPP took 1.75 million events using 120~GeV/c primary beam protons
impinging on the NuMI (spare) target. These events will play a crucial
role in the prediction of neutrino fluxes in the NuMI beamline and
will enable the MINOS experiment to control the systematics in the
near/far detector ratios as well as helping them understand the near
detector performance. Figure~\ref{minos} shows a radiograph of the
MIPP measurements of the MINOS target.  
The graphite slabs and cooling tubes can be seen.
These events were obtained during the commissioning phase of this 
target measurement where the beam was not yet fully focused and 
aligned on the target. The 1.75~Million events on the NuMI target were 
obtained after the beam was aligned and centered on the target. 
Figure~\ref{ringrads} shows the rich ring radii vs momentum of positive 
tracks originating from the NuMI target. Superimposed are the curves for 
known particles. This shows the excellent particle identification of the
 MIPP detector for forward going particles.

%% file: atmospheric.tex
\subsection{Particle production on Nitrogen and the question of Cosmic Ray  Showers in the Atmosphere}

With the upgraded MIPP, we propose to measure particle production on a cryogenic nitrogen
target using positive and negative beams, which is needed by experiments
measuring cosmic ray air showers (Pierre Auger, HiRes etc) and also
atmospheric neutrinos (Amanda, Ice Cube, HyperK etc). The problem is
illustrated in a recent paper~\cite{meurer} which simulates the air
showers produced by protons of $10^6$~GeV energy in the
atmosphere. The shower goes through several generations of
interactions and produce pions and kaons that decay to produce muons
and neutrinos. The muons and neutrinos are observed in the detectors
and are termed the daughter particles. The mesons that produced the
muons and neutrinos are termed the mother particles and the particles
that interacted in the atmosphere to produce the mother particles are
termed the grandmother particles in the jargon.

\begin{figure}[htb!]
\begin{minipage}{15pc}
\includegraphics[width=\textwidth]{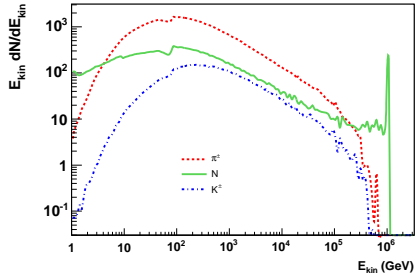}
\end{minipage}
 \caption{The energy distribution of the grandmother particles in a
 vertical air shower produced by $10^6$~GeV proton interacting in the
 atmosphere as a function of particle type. These particles interact
 further in the atmosphere to produce more particles which then decay
 into muons and neutrinos. The muons are detected at a distance of 
 0-500 meters from the shower center at 
 ground level. It can be seen that these spectra peak at
  energies relevant to the MIPP energy scale. }
\label{meurer1}
\end{figure}

\begin{figure}[htb!]
\begin{minipage}{15pc}
\includegraphics[width=\textwidth]{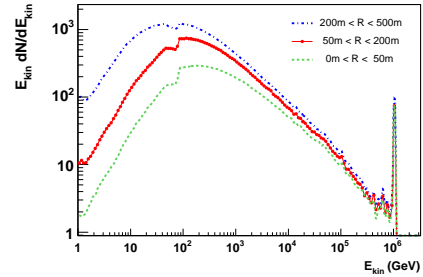}
\end{minipage}
 \caption{The energy distribution of the grandmother particles in a
 vertical air shower produced by $10^6$~GeV proton interacting in the
 atmosphere as a function of distance from the shower center. 
The spectrum peaks 
at energies relevant to the MIPP energy scale.}

\label{meurer2}
\end{figure}

Figure~\ref{meurer1} shows the energy spectrum of the grandmother
particles ($\pi, K$ and $p$ in an air shower that are produced
by a primary proton of $10^6$~GeV. The spectrum for the pions peaks at
100~GeV and the kaons and protons at somewhat higher energies. These
particles interact with the nitrogen (and oxygen) in the atmosphere to
produce the atmospheric neutrinos and muons. In other words, the beam
energies available at MIPP are relevant to the simulation of the
cosmic ray air showers. The muon flux measurement is a critical component 
of estimating the energy scale of the cosmic ray shower. MIPP measurements thus
will help reduce the systematics in the cosmic ray energy scale measurements. 
As the primary cosmic ray energy increases, the peaks in this plot do not 
shift to higher energies. 
Understanding the shower systematics at the peak of this 
spectrum (i.e MIPP energies) will help the energy systematics of 
cosmic rays of all energies.
Figure~\ref{meurer2} shows the distribution of grandmother particles
at different lateral distances from the shower center for all particle
types.  For the lower energy interactions, the simulation code Gheisha
is used to simulate the interactions of the particles with the
atmosphere. For higher energy interactions, the simulation code QGSJET 01
is used. The sharp break in the spectra at 100~GeV is where the two
codes meet and disagree at places by a factor of two. This illustrates
the problem. These codes at present are ``tuned'' on single arm spectrometer
data and disagree with each other.

Most of the data acquired to date of relevance to air showers
are over 25 years old and were obtained using Beryllium targets and
single arm spectrometers resulting in a discrete angular
coverage. MIPP will measure the outgoing pion and kaon spectrum for
proton and pion beams in its full secondary beam momentum range. These
cross sections are the most important in simulating cosmic ray showers
in the atmosphere.  In addition, it will also have kaon and antiproton
interactions on nitrogen of which virtually nothing is known. Being an
open geometry experiment, the MIPP angular coverage will be
continuous, not discrete.

The need for MIPP data is recognized by the cosmic ray community, 
some of whom have joined the experiment.
\subsubsection{MIPP Measurement of $\pi/K$ ratios}

Because of its excellent particle identification capabilities, MIPP
upgrade will measure the ratio of charged kaons to pions as a function
of $p_L,p_T$ of the final state particle. This measurement is of
importance to both the NuMI target measurements and the atmospheric
neutrino measurements, since the charged K's produce $\nu_e's$ which
are a background to the oscillation search $\nu_\mu\rightarrow \nu_e$.

%% file: run_plan.tex
\section{MIPP Upgrade Proposal}
In October 2006, MIPP proposed to upgrade its detector by increasing
the data acquisition speed of its TPC from 20~Hz to 3000~Hz using the
ALTRO/PASA chips developed for the ALICE
collaboration~\cite{upgr}. With this large factor in data acquisition
speed, it becomes possible to acquire 5~million events in a single day
of running. The electronics of the rest of the detector will also be
upgraded to run at this speed. We propose to use a hemisphere of the
plastic ball detector to measure the target recoil particles. This
enables us to measure particle production on 30 of the most common
nuclei found in particle physics detectors and improve the hadronic
shower simulator programs. It will also permit us to study
non-perturbative QCD in unprecedented detail. The baryon resonance
spectrum can be investigated up to 3~GeV/c$^2$, using pion and kaon
beams in the momentum range 1~GeV/c-5~GeV/c. The decision on the
upgrade proposal has been deferred till analysis results from the
present run are available and the collaboration has been strengthened
further.

\subsection{Upgrade Run Plan}

We propose to conduct the upgrade running in three phases. In the
first phase, we will acquire data on the neutrino targets, liquid
nitrogen and take particle production measurements on 12 thin nuclear
targets (5 million events per nucleus) which will improve the quality
of hadronic shower simulation programs. The number of events acquired
in each target is indicated in table~\ref{phase1}.
\begin{table}[tbh]
\begin{minipage}{15pc}
\begin{tabular}{ccc}
\hline
Target & Events & Running Time  \\
       & (Millions)       &  (Days)     \\
\hline
NuMI target 1 & 10 & 2  \\
NuMI target 2 & 10 & 2 \\
Liquid Hydrogen & 20 & 4  \\
Liquid Nitrogen & 10 & 2  \\
12 Nuclei & & \\
$D_2$,Be,C,Al,Si,Hg, & & \\
Fe,Ni,Cu,Zn,W,Pb & 60 & 12  \\
Total Events & 110 & 22  \\
\hline
\end{tabular}
\caption{Phase 1 Run Plan.}
\label{phase1}
\end{minipage}
\end{table}
\begin{table}[tbh]
\begin{minipage}{15pc}
\begin{tabular}{ccc}
\hline
Target & Events & Running Time  \\
       & (Millions)       &  (Days)     \\
\hline
18 Nuclei & & \\
Li,B,$O_2$,Mg,P, & & \\
S,Ar,K,Ca,Ni,Nb,Ag, & & \\
Sn,Pt,Au,Pb,Bi,U & 90 & 18  \\
10 Nuclei B-list & &   \\
Na,Ti,V, Cr,Mn,Mo,& & \\
I, Cd, Cs, Ba & 50 & 10  \\
Total Events & 140 & 28  \\
\hline
\end{tabular}
\caption{Phase 2 Run Plan.}
\label{phase2}
\end{minipage}
\end{table}

During phase 2, we plan to complete the remaining 18 nuclei of the
A-List, as detailed in the section on hadronic shower simulation and then
proceed with the B-list if there is need. The second phase of running
is detailed in table~\ref{phase2}.

During phase 3, we plan to go into the tagged neutral beam mode, where we
run the liquid hydrogen target and allow the ILC calorimetry to run
simultaneously in place of the MIPP calorimeter to study the response of the 
ILC calorimeters to tagged neutral beams.
\section{Conclusions}
MIPP has acquired high quality particle production data which it is
busy analyzing. The MIPP upgrade will improve quality and statistics
data by an order of magnitude by speeding up the
upgrade. Collaborators are welcome.

%% file: raja_isvhecri06.bbl
\begin{thebibliography}{9}
\bibitem{e690}
The multi-cell \v Cerenkov detector was  first used in 
BNL experiment E766, then in  Fermilab
experiment E690  followed by other BNL experiments.
\bibitem{ecal}The EM calorimeter was fabricated
by MIPP and uses lead as the radiator and an array of proportional
tubes with 2.54 cm wire spacing as the readout.  It has ~10 radiation
lengths and has 10 longitudinal segments.
\bibitem{hcal} Theiron/scintillator fiber
hadron calorimeter is recycled from the
HyperCP collaboration. It has 9.7 interaction lengths and has 4 longitudinal
and 2 transverse segmentations.
\end{thebibliography}
